**Allovalency revisited: an analysis of multisite phosphorylation and substrate rebinding**

Jason W. Locasale[1]*

[1]Department of Biological Engineering, Massachusetts Institute of Technology, 77

Massachusetts Ave.  Cambridge, MA 02139. *Locasale@MIT.edu

## Abstract

The utilization of multiple phosphorylation sites in regulating a biological response is ubiquitous in cell signaling. If each site contributes an additional, equivalent binding site, then one consequence of an increase in the number of phosphorylations may be to increase the probability that, upon disassociation, a ligand immediately rebinds to its receptor. How such effects may influence cell signaling systems has been less studied. Here, a self-consistent integral equation formalism for ligand rebinding, in conjunction with Monte Carlo simulations, is employed to further investigate the effects of multiple, equivalent binding sites on shaping biological responses. Multiple regimes that characterize qualitatively different physics due to the differential prevalence of rebinding effects are predicted. Calculations suggest that when ligand rebinding contributes significantly to the dose response, a purely allovalent model can influence the binding curves nonlinearly. The model also predicts that ligand rebinding in itself appears insufficient to generative a highly cooperative biological response.

## INTRODUCTION

The establishment of precise controls within signaling modules is an evolutionary prerequisite for a robustly functioning cellular system. A central issue to such control is the regulation of a dose response or the necessary input-output relationships that direct a specific biological function[1,2]. One such input that is widely utilized in many biological systems is the number of phosphorylations on a protein containing many potential phosphorylation sites. Multisite phosphorylation is ubiquitous in cell biology and regulates myriad cell decisions[3-5].

One salient example comes from the regulation of the cell cycle by ubiquitin mediated protein degradation, a key motif in the control of the cell cycle[5,6]. In the seminal work by Nash et al.[7], the authors show that the CDK inhibitor, Sic1 functions through a thresholding mechanism – Sic1 must be phosphorylated at least 6 six (of its 9 possible) sites in order to be ubiquitinated and subsequently targeted for degradation. Sic1 is intrinsically disordered[8] and the location and specificity of these six phosphorylation sites seems to be unimportant at least to some extent. This observation among others[9] led to the hypothesis that the function of these seemingly redundant post translational modifications may be to increase the probability that Sic1 rebinds to its substrate upon disassociation[10,11] and a mathematical model[10] was developed to investigate the rebinding of a polyvalent ligand. In this model, a ligand, once disassociated, effectively escapes from its receptor unless it is phosphorylated a sufficient number of times so as to increase its chances of rebinding.

The problem of ligand rebinding has been extensively studied in many contexts[12-17]. Some of the most comprehensive studies were carried out in the context of two settings: 1.) ligand binding/unbinding to and from a surface as a model for the kinetics of ligand binding to cell-surface receptors[13,15,18] and 2.) chemotaxis and autocrine signaling resulting in rebinding of a ligand secreted from a cell[12,19,20]. In each of these studies, it was demonstrated that ligand rebinding can be very significant. Despite these advances, how changes in the phosphorylation state of a substrate is related to rebinding and how this affects a biological dose response curve has not been fully investigated. A schematic of this effect is shown in Fig. 1.

Towards this end, we use an integral equation theory and Monte Carlo simulations to study the rebinding of a ligand to a receptor from which it initially disassociated and how this rebinding may be affected by multiple recognition sites. From considering only the effects of a single molecule rebinding to its receptor, we compute the time dependence of the probability that a ligand remains bound as a function of the number of phosphorylations. In turn, we compute the probability that a ligand escapes its target as function of the number of recognition sites. The model and numerical simulations predict that this escape probability can decrease nearly exponentially as a function of the number of independent binding sites thus suggesting that ligand rebinding greatly affects the binding kinetics. We also highlight the importance of two physical regimes of ligand rebinding that are characterized by weak and strong rebinding and show how each regime may affect the input-output relationships of a system with multiple phosphorylation sites. We further note that the model predicts that, although a ligand's propensity to immediately rebind, as a function of the number of available binding sites, greatly affects the shape of the biological response, additional mechanistic ingredients appear to be required to achieve a highly cooperative response. Finally, we note that while our model predicts that the probability of a polyvalent ligand escaping from its receptor decreases exponentially as a function of the number of binding sites, this property appears insufficient to give rise to a highly cooperative response as has been previous predicted[10]. The source of this discrepancy appears to lie in how the rate constants in the previous phenomenological model were varied independently to achieve the desired cooperativity.

## METHODS AND MODEL DEVELOPMENT

### Multisite phosphorylation and ligand rebinding

The key considerations that are used to develop our model lie in the questions that we wish to address in this study. In particular, our aim is to investigate how ligand rebinding may be affected by multisite phosphorylation. Other studies of multisite phosphorylation have investigated the consequences of other physical effects such as distributive phosphorylation and feedback regulation[4,21]. We are interested in computing the probability that a ligand remains bound as a function of time and as a function of the number of recognition sites on the receptor.

To model this scenario, we assume that at time zero, a ligand is bound to its receptor and can be released with a constant unit time probability. When the ligand is in immediate proximity of the receptor, there is a probability $\theta$ that the ligand rebinds to the receptor within the time it takes to diffuse away from the immediate vicinity of the recognition domain. Multiple phosphorylations are then parameterized by a change in this probability. In the case we consider, which we refer to as the 'allovalent' model[10], each phosphorylation contributes equally and independently to the value of the parameter $\theta$; i.e.

$$\theta = n\theta_0,$$

where $n$ is the number of phosphorylations and $\theta_0$ is the probability that a ligand that is proximally located to the recognition site will rebind when it is singly phosphorylated on any site. In the work by Klein et al[10], this assumption (that each site contributes equally and independently to the rebinding probability) was sufficient to give rise to a highly cooperative response. Our aim is to further investigate the consequences of such an assumption.

Important to note is that in order for $\theta$ to be a probability it must be less than or equal to one. Therefore, $\theta_0$ must be bounded by $\dfrac{1}{N}$;

$$\theta_0 \leq \frac{1}{N},$$

where $N$ is the maximum number of phosphorylation sites on the ligand. An additional complication that is not considered here is the time dependence of $n$ that may become important at late times. The theory therefore aims to investigate solely how rebinding is affected given a fixed number of binding sites. Also, this description of ligand binding is considered to be a "mean field" treatment since all conformational fluctuations of both the ligand and its receptor are neglected by the introduction of the parameter $\theta$. One could also imagine that $\theta$ could have a complex, nonlinear dependence on $n$ for a given $\theta_0$ (i.e. $\theta = f\left(n;\theta_0\right)$) as would be the case when cooperative electrostatic interactions among the multiple phosphate groups influence binding[22].

**A self-consistent integral equation theory for ligand rebinding**

To begin our analysis, we exploit a formalism that monitors the trajectories of a single ligand as it disassociates from and potentially rebinds to its target to which it was originally bound. The formalism was developed by Tauber et al.[15] who investigated the effects of a ligand binding to a collection of receptors on a planar surface in the context of surface plasmon resonance studies. Although our approach is similar in many regards, there are some subtleties that distinguish the two approaches and they are discussed within our treatment. We consider an equation that describes the time-evolution of the probability, $f(t)$, for a single ligand to be bound to its receptor provided that it is initially bound to its target. The initial condition $f(0) = 1$ is used. Single molecule master equations of this sort have been used extensively in many different contexts[23].

A knowledge of this function allows one to compute the probability that a ligand is bound as a function of time as well a time dependent escape probability which is taken to be, $1 - f(t)$. A differential equation for the time evolution of $f(t)$ can be written as

$$\frac{df}{dt} = \upsilon_+ - \upsilon_- \qquad (1)$$

The negative contribution $\upsilon_-$ simply follows first order disassociation kinetics (i.e. $\upsilon_- = k_- f(t)$). Thus, in the absence of rebinding ($\upsilon_+ = 0$), $f(t)$ decays via a single exponential with time constant $\frac{1}{k_-}$, $f(t) = e^{-k_- t}$. $\upsilon_+$ on the other hand is entirely due to the contribution from the rebinding of a single previously disassociated ligand. The forward rate of binding, $\upsilon_+$, is therefore the probability that a protein dissociates in the interval $\tau$ and $\tau + d\tau$ and then subsequently rebinds at a later time interval, $t - \tau$ and $(t - \tau) + d\tau$, integrated over all previous times, $\tau$. An equation for $\upsilon_+$, therefore, can be written as follows:

$$\upsilon_+ = k_- \int\limits_0^t d\tau f\left(\tau\right) R\left(\delta, t-\tau\right). \qquad (2)$$

$R\left(\delta, t'\right)$ is the probability per unit time that a protein binds to its target in the time interval $\{t', t'+dt\}$ given that it is located a distance, $\delta$, away from the target at time 0 ($\delta$ is the small distance from the receptor that the ligand is placed when it disassociates).

Combining eqs. 1 and 2, we obtain an integral equation that accounts for the state of the ligand as a function of its entire history:

$$\frac{df\left(t\right)}{dt} = k_- \left[ \left\{ \int\limits_0^t d\tau f\left(\tau\right) R\left(\delta, t-\tau\right) \right\} - f\left(t\right) \right]. \qquad (3)$$

We can analyze eq. 3 first by introducing Laplace-transformed variables:

$$\tilde{f}\left(s\right) = \int\limits_0^\infty dt e^{-st} f\left(t\right) \text{ and } \tilde{R}\left(\delta, s\right) = \int\limits_0^\infty dt e^{-st} R\left(\delta, t\right).$$

By substituting the Laplace transforms into eq. 3 and making use of the convolution theorem[24], we obtain:

$$s\tilde{f}\left(s\right) - f(0) = k_- \tilde{f}\left(s\right) \left[ R\left(\delta, s\right) - 1 \right]. \qquad (4)$$

Or, upon rearranging and inverting the Laplace transform:

$$\tilde{f}\left(s\right) = \frac{f\left(0\right)}{s + k_- \left[ 1 - R\left(\delta, s\right) \right]}$$

$$f\left(t\right) = \frac{1}{2\pi i} \int\limits_{c-i\infty}^{c+i\infty} ds e^{st} \frac{f\left(0\right)}{s + k_- \left[ 1 - R\left(\delta, s\right) \right]}. \qquad (5)$$

Thus, the probability that a protein remains bound can be solved exactly provided that an explicit form of $R(\delta, s)$ can be obtained and that the resulting Laplace inversion can be computed.

A convenient way to obtain $R(\delta, s)$ as developed previously[15] is to compute the quantity self-consistently by considering the statistics of first passage processes for an individual protein disassociating from its ligand: i.e,

$$R(\delta, t) = \theta F(\delta, t) + (1 - \theta) \int_0^t d\tau R(\delta, t - \tau) F(\delta, \tau) \qquad (6)$$

where $\theta$ is a parameter that gives the probability that the protein will bind to its substrate given that it is within a distance $\delta$; and, $F(\delta, t')dt'$ is the probability that a protein first reaches the origin, starting from a distance $\delta$ at time 0, in the time interval $\{t', t' + dt'\}$. In the case we study, $\theta$ is a linear function of the number of phosphorylations $n$, $\theta = n\theta_0$ where $\theta_0$ is the probability that a ligand binds given that it has been singly phosphorylated. The contribution of the first term in eq. 6 is from the probability that a ligand is absorbed the first time it reaches its target. The contribution of the second term is from the probability that the ligand reached the target at time $\{\tau, \tau + dt\}$, was reflected (i.e. the ligand did not bind before it diffused away) at that time, and was then later absorbed at $\{t - \tau, (t - \tau) + dt\}$.

Again, upon Laplace transforming eq. 6 and the first passage time PDF, i.e.

$\tilde{F}(s) = \int_0^\infty dt e^{-st} F(\delta, t)$, and again, noting the convolution theorem, eq. 6 becomes:

$$\tilde{R}(\delta,s) = \theta \tilde{F}(\delta,s) + (1-\theta)\tilde{R}(\delta,s)\tilde{F}(\delta,s)$$

$$\tilde{R}(\delta,s) = \frac{\theta \tilde{F}(\delta,s)}{1-(1-\theta)\tilde{F}(\delta,s)}. \tag{7}$$

In the work by Tauber and coworkers[15], a similar equation as eq. 7 is used to compute $R(\delta,t)$. However, in their treatment of the calculation of $R(\delta,t)$, the coefficients $\theta$ and $1-\theta$ in eq. 6 are replaced with $\theta(1-f(t))$ and $1-\theta(1-f(t))$. In their problem, the authors considered rebinding to receptors on a planar surface and the probability that a ligand reaches a receptor that contains a ligand that already contains a bound ligand need be taken into account. Our equation for the absorption probability does not require the additional $1-f(t)$ factor since we are only considering the rebinding of a receptor to a single isolated receptor.

For further analysis, the first-passage time distribution function is now required and is considered in three dimensions. Assuming spherical symmetry, the solution to the first passage problem can be obtained in the Laplace domain and its derivation is contained in the appendix; thus,

$$\tilde{F}(a+\varepsilon;s) \approx \left[\frac{a}{a+\varepsilon}e^{-\sqrt{\tau s}}\right]. \tag{8}$$

The distance, $\delta$, is written as $\delta = a+\varepsilon$ ($a$ is the radius of the recognition domain and $\varepsilon$ is the average distance away from the boundary of the recognition domain that the ligand is initially displaced when it rebinds) the variable $\tau = \frac{(a+\varepsilon)^2}{D}$ has been introduced along with $D$ being the diffusion constant of the ligand. $\tau$ is the time it takes for the ligand to diffuse a distance on the order of the distance to its target.

A further simplification can be made if we observe the system on time scales commensurate with the disassociation time; $t \sim 1/k_-$, i.e. $t \gg \tau$ (so that s is small).

$\tilde{F}(a+\varepsilon;s)$ becomes:

$$\tilde{F}(a+\varepsilon,s) \approx \frac{a}{a+\varepsilon}\left[1-\sqrt{\tau s}\right] + O(\tau s). \qquad (9)$$

This approximation has been shown to be very good in one dimension[15] in which rebinding is believed to be more prominent. Therefore, up to order $O(\tau s)$, we substitute eq. 9 into eq. 7 and obtain:

$$1-\tilde{R}(a+\varepsilon,s) \approx \frac{(1-\gamma)+\gamma\sqrt{\tau s}}{1-(1-\theta)\gamma} \qquad (10)$$

where $\gamma = \frac{a}{a+\varepsilon}$.

Inserting this expression into the integrand in eq. 5 yields:

$$\tilde{f}(s) = \frac{f(0)}{s+k^{eff}_-\left[(1-\gamma)+\gamma\sqrt{\tau s}\right]}$$

$$f(t) = \frac{1}{2\pi i}\int_{c-i\infty}^{c+i\infty} ds \frac{e^{st}f(0)}{s+k^{eff}_-\left[(1-\gamma)+\gamma\sqrt{\tau s}\right]} \qquad (11)$$

where, $k^{eff}_- = \frac{k_-}{1-(1-\theta)\gamma}$.

**Monte Carlo simulations**

To supplement the theory, we also considered Monte Carlo simulations. Simulations were performed by considering a collection of random-walkers with a set of receptors on a three-dimensional lattice of 100 x 100 x 100 lattice spacings. Each protein (receptor and ligand) occupy one site on the lattice at any given time. In each Monte

Carlo step, with equal probability for a move to be made in any direction, an attempt to allow a molecule to diffuse is given by $P_{diff}$ which defines a time sale that then defines a diffusion constant; i.e. $P_{diff} \sim D / L^2$ where D is the diffusion constant and L is the length of a lattice spacing which is taken to be the diameter of a typical protein or in this case, $L \sim 10nm$. When encountering an immobile receptor at any of its nearest-neighbor positions, the substrate can bind with probability $P = P_{rxn} e^{-\left(\frac{E_{k_+}}{k_b T}\right)}$, so that $k_+ \sim e^{-\left(\frac{E_{k_+}}{k_b T}\right)}$. $k_b T$ is Boltzman's thermal energy, $E_{k_+}$ is the energy barrier for association when a receptor and ligand come into contact. In this scheme the rebinding probability $\theta$ behaves as, $\theta = n\theta_0 \sim e^{-\left(\frac{E_{k_+}}{k_b T}\right)}$.

The fraction of bound ligands was computed by sampling at steady-state, as a function of $\theta$, $\theta \propto k_+$. Escape probabilities were computed by first allowing a receptor to release its ligand at time $t = 0$; at a later time, $t = t_0$, sampling of whether or not the ligand is again bound to its target is performed. $t_0$ was chosen to be a time on the order of the encounter time for a protein in a eukaryotic cell; $t_0 = 1000 mcsteps$ ($1000 mcsteps \sim 1ms$ assuming a lattice spacing of $L = 10nm$ and a diffusion constant $D = 10\mu m^2 / s$). For each value of $\theta$, the statistics determining the escape probability were obtained from 100,000 independent trials.

## RESULTS AND DISCUSSION

### Rebinding probabilities

From eq. 11, the relevant biological quantities can be computed. First consider the absorption probability in the Laplace domain. A numerical inversion of eq. 11 can in principle be accomplished and the subsequent function plotted. However, since such a computation is difficult to accomplish due to numerical instabilities resulting from the multi-scale nature of the computation, we considered the function in the Laplace domain. By substituting the results contained in eq. 8 into the expression for $R(\delta, s)$ (eq. 7), we obtain.

$$\tilde{R}(\delta, s) = \frac{\theta \gamma e^{-\sqrt{\tau s}}}{1 - (1 - \theta) \gamma e^{-\sqrt{\tau s}}} \qquad (12)$$

As seen in Fig. 2a, since the first-passage time distribution decays as a stretched exponential function in the Laplace domain, rebinding can be significant over many time scales.

**Kinetics of disassociation modified by rebinding events—exponential versus non-exponential decay giving rise to 'strong' and 'weak' regimes of rebinding**

In one dimension, for all parameter ranges, rebinding events lead to strongly non-exponential kinetics whenever significant rebinding is possible (Appendix). That is, as a result of rebinding, a ligand can remain bound to its receptor long after the time scale that characterizes its dissociation. In three dimensions, the effects of rebinding should be less significant since fewer returns to the origin occur in higher dimensions and some trajectories never return to the origin[25].

Upon inspection of the Laplace inversion in eq. 12, two kinetic regimes are observed that depend on the relative size of the receptor as determined by

$\gamma = \dfrac{a}{a + \varepsilon}$ . First, if

$$(1-\gamma) << \gamma\sqrt{\tau s} \qquad (13)$$

(e.g. the radius of gyration of the disordered protein is small compared to the radius of the region to which it binds to its targeted substrate, $\varepsilon \approx 0$ and $\gamma \sim 1$), then the overall kinetics of ligand disassociation that are modified as a result of rebinding events behave in a similar fashion to that of the one-dimensional case[15] as shown in the appendix. This can be seen by taking the $\varepsilon \to 0$ (i.e. $\gamma \to 1$) limit of eq. 11 in which case,

$$f(t) \to f(0)e^{\kappa t}erfc\left(\sqrt{\kappa t}\right) \qquad (14)$$

where, $\kappa^{-1}$ is a time scale that behaves as $\kappa \to \dfrac{4k_-^2\alpha}{\theta^2}$ as $\gamma \to 1$ (appendix).

One the other hand, for $\gamma$ significantly less than one, $(1-\gamma) >> \gamma\sqrt{\tau s}$, since $\tau$ is a microscopic time scale, an exponential decay is observed:

$$f(t) \sim e^{-k_d t} \qquad (15)$$

where $k_d = \dfrac{k_-(1-\gamma)}{\left[1-\gamma(1-\theta)\right]}$.

In Fig. 2b, plots of the decay of the probability $f(t)$ are shown for three cases. In the first case, no rebinding binding ($\theta = 0$) is considered and $f(t)$ behaves according to: $f(t) = f(0)e^{-k_- t}$. In the second case, strong rebinding is considered ($\gamma \to 1$) so that $f(t)$ takes on highly non-exponential behavior; i.e. $f(t) = f(0)e^{\kappa t}erfc\left(\sqrt{\kappa t}\right)$. Finally, in the third case, weak rebinding is considered ($\gamma < 1$) so that $f(t)$ takes the form: $f(t) = f(0)e^{-k_d t}$.

The parameter values used are given in the figure caption. As shown in the plot, the two regimes of rebinding lead to dramatically different consequences. When the ligand begins significantly far away from its target leading to the weak ($\gamma < 1$) rebinding regime, rebinding serves simply to decrease the off rate ($k_d < k_-$). In contrast, when the ligand begins close to its target (or the target is very large in comparison to the ligand), $\gamma \to 1$ and the shape of the disassociation curve changes dramatically resulting in nonexponential disassociation kinetics. The presence of a distribution with a fat tail (i.e. $\sim t^{-1/2}$; $t >> (1/\kappa)$) is observed; this signifies that the release of the ligand is distributed over many time scales – the ligand becomes trapped by the receptor for long times.

**The fraction of bound ligands can be greatly influenced by rebinding**

With the formulas obtained in eqs. 14 and 15, $f(\theta, t_0)$, the probability that a ligand remains bound as a function of $\theta$, can be studied at different time points, $t_0$. Shown in Figs. 3a and 3b, the behavior of these functions is plotted. For the strong rebinding ($\gamma \to 1$) case in fig 3a, it can be seen that the fraction of bound ligands is strongly influenced by rebinding over a broad range of time scales (i.e. 0.001s – 1000s). On the other hand, for weak rebinding, the fraction of bound ligands is only strongly influenced by rebinding on a time scale, $\tau_-$ commensurate with the intrinsic off-rate (i.e. $\tau_- \sim \dfrac{1}{k_-}$). Such behavior is a direct consequence of the non-exponential vs. exponential shapes of the decay curves. It is also noted that fitting each curve to a Hill function $\dfrac{\theta^H}{K_{50\%}^H + \theta^H}$ by nonlinear regression, gives a value of $H \sim 1$ for all curves indicating a 'Michaelian' dose response[2].

**Escape probabilities and the effects of rebinding on dose response curves**

The escape probability can be computed within this theory from a consideration of the fraction of bound ligands $f(\theta, t_0)$. $1 - f(\theta, t_0)$ gives the probability that a ligand is not bound to its target at time $t_0$ (i.e. the probability that the ligand has "escaped"). As seen in the plots in Figs. 3a and 3b, for large enough values of $\theta$, long after the disassociation from the first order decay process, ligands can be trapped by their receptors.

In the weak rebinding regime, the escape probability has the functional form:

$$1 - f(\theta, t_0) \approx 1 - \exp\left(-\frac{a}{[b + c\theta]} t_0\right)$$ as can be seen upon rearranging eq. 15. On the other hand, in the strong rebinding regime (eq. 14), the escape probability behaves as:

$$1 - f(\theta, t_0) \approx 1 - e^{a t_0 \theta^{-2}} erfc\left(a\theta\sqrt{t_0}\right).$$ For typical parameter values, these functions decay at rates commensurate with the rates of an exponential process characterized by a single time scale as seen in Figs. 3a and 3b.

Alternatively, Monte Carlo simulations[26] can be used to compute the escape probabilities numerically. Plots of the escape probabilities are shown in Fig. 4a; as indicated on the inset of the plot, the data obtained from the Monte Carlo simulations are shown to fit well to an exponential decay function with a single parameter i.e. $f(\theta, t_0) \sim e^{-k\theta}$. The fraction of receptors bound as a function of $\theta$ is also computed from the computer simulations and plotted in Fig. 4b. Different values of receptor density are considered. For each curve, as exemplified on in the inset of Fig. 4b, a fit to a Hill function gives a Hill coefficient of near unity. The plots in Fig. 4b. are consistent with those obtained from the theory and plotted in Fig. 3a.

While the curves in Figs. 3a,b and 4b show that the Hill coefficient is near unity, thresholding effects in the dose response curves may appear when rebinding is significant. This thresholding effect that is observed in Figs. 3a, b, and 4b is defined as a different value of $\theta$ needed to reach a given value of $f\left(\theta, t_0\right)$. These results are thus similar to the observations that have been previously reported[4] that considered the case of multiple phosphorylation steps that occur in an ordered, distributive manner. This result is therefore expected to become more prominent upon incorporation of the possibility of rebinding.

Finally, we considered how the fraction or probability that a ligand remains bound vary as a function for the number of phosphorylations, $n$ for different values of $\theta_0$ (recall: $\theta = n\theta_0$). Four cases are shown: the strong rebinding ($\gamma \to 1$) case at long ($100s$) and short ($10s$) times (Figs. 5a,b), and the weak rebinding ($\gamma < 1$) at long ($5s$) and short ($1s$) times (Figs. 5c,d). As seen, graded responses are observed in each of these cases. Perhaps interesting to note is the non-uniformity of these dose response curves; some appear near linear while others have a nonlinear, hyperbolic shape. This effects results from a rescaling of $\theta_0$. For different values of $\theta_0$, the response to changing values of $n$ is different.

## Comparison to previous theoretical work on ligand rebinding and multisite phosphorylation

Previous theoretical work has also studied the effects of multisite phosphorylation. In a prior study[10], a theoretical model predicted an exponential decay in the escape probability as a function of the number of phosphorylations. This exponential decay was predicted to be sufficient to give a highly cooperative dose response curve (the addition of a single independent binding site results in a large increase in the fraction of ligands bound to their receptors). The model that was developed consists of a ligand existing in one of three states: bound to the receptor (B), in a region proximal to the receptor (P), and a region far away from the receptor (F). Transitions between these states are considered that result in the following kinetic scheme with kinetic constants, $k_{on}, k_{off}, k_{esc}, k_{cap}$;

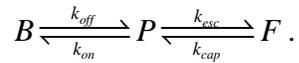

The fraction of bound ligands is taken to be the proportion of ligands in the bound and proximal states (i.e. $f_B = \dfrac{B+P}{L_{tot}}$ where $B$ is the number of ligands in the bound state, $P$ is the number of ligands in the proximal state, and $L_{tot}$ is the total number of ligands.). Using mass action kinetics, an expression for $f_B$ as a function of the kinetic constants and number of available receptors $R_f$ can be computed[10]:

$$f_B = \frac{1}{1+\kappa} \qquad (16)$$

where,

$$\kappa = \frac{k_{esc}}{k_{cap}} \left[ R_f \left( 1 + \frac{k_{on}}{k_{off}} \right) \right]^{-1} . \qquad (17)$$

$k_{off}$ is an off rate for dissociation of a bound ligand and $k_{on}$ is an on rate that is taken to be proportional to the number of phosphorylations ($k_{on} \propto n$). $k_{esc}$ is computed from first passage time statistics and is shown to decay exponentially $k_{esc} \propto e^{-an}$ as a function of the number of phosphorylations $n$. It is also assumed that $k_{cap}$ is a diffusion limited rate constant[17] and is taken to be independent of $n$[10]; that is, $k_{cap} = 4\pi R_0 \left( D_l + D_r \right) N_A$ where $R_0$ is the radius of the receptor, $D_l$ and $D_r$ are diffusion constants for the ligand and receptor respectively, and $N_A$ is Avogadro's number.

From the expression, it is clear that, for some parameter values, a highly cooperative response[2] can obtained when $k_{esc}$ decays exponentially for increasing $n$ while keeping $k_{cap}$ fixed. However, in the framework of the model, it is not clear why $k_{cap}$ (a rate constant of diffusion limited capture for ligands to enter the proximal region near the receptor) is independent of $n$ or $k_{esc}$. If more ligands are immediately rebinding to their receptors and as a result $k_{esc}$ decreases, then fewer ligands are available to diffuse into the proximal region denoted by the $F \rightarrow P$ transition. $k_{cap}$, it seems, should decrease accordingly.

Consistent with this model, the theory and Monte Carlo simulations both predict a fast (and nearly exponential) decrease in the probability of escape for a ligand as a function of the number of phosphorylations as seen in Fig. 4a. This exponential decrease in the probability of escape of a newly disassociated ligand, however, appears insufficient to produce a highly cooperative response due to rebinding (Fig. 4b). The discrepancy between these two findings appears to lie in the assumption of a constant value of $k_{cap}$ that was used. The Monte Carlo simulations show that the allovalent model predicts a Michaelian[2] (i.e. Hill coefficient of unity) response. This result is perhaps not surprising since there is no cooperativity introduced into the model. The effect of rebinding, in itself, appears insufficient to give a cooperative response. However, despite this apparent lack of cooperativity, differential rebinding effects (with respect to changes in the number of binding sites) can be very significant as has been emphasized throughout this work.

## Summary

We first reformulated the problem of the rebinding of a protein with multiple independent phosphorylation sites, to its target in the context of a self consistent integral equation theory[14,15], to study the effects of one dimensional ligand rebinding to a surface containing antibody receptors. Within this formalism, we solved the rebinding problem of a single ligand to an isolated receptor in three dimensions in two limits that depend on the relative sizes of the receptor and ligand. We find two qualitatively distinct regimes of rebinding kinetics whose crossover depends mainly on the size of the substrate and its target. In one regime (i.e. when there is strong rebinding), the kinetics of ligand disassociation takes on a similar functional form to that of the one-dimensional case— this results resulting in a slow decay of bound substrates characterized by non-exponential kinetics and a power law tail. Alternatively, in the other regime (i.e. when there is weak rebinding), the behavior of the kinetics of disassociation exhibits an exponential form and is thus characterized by a single rate constant – rebinding gives simply a slower time constant signifying a lesser influence on rebinding. The model predicts that the relative size of the ligand (that determines the rebinding regime) may play a key role in determining the functional role of multisite phosphorylations. It may be interesting to study how the different regimes of rebinding, that are predicted in this model, relate to other biological processes that require ligand rebinding at different length and time scales, such as autocrine signaling[27].

We then used the results obtained to compute rebinding probabilities. We showed that, in some instances, rebinding can occur over many time scales and contribute significantly to the total bound fraction of ligands. Furthermore within this model, an increase in the number of independently acting phosphorylation sites leads to a near exponential decrease in the probability that a ligand escapes from its target (i.e. it diffuses a large distance without being captured by its target). The model also predicts a graded response[2] and yields a Hill coefficient of near unity for all parameter values. Thus, statistically independent contributions to the association rate of the ligand in the form of additional binding sites and their additive effect on the association rate (while potentially having a great impact on the binding kinetics) does not appear to in itself yield a highly cooperative response. These additional binding sites can, however, influence the shape of the dose response in a nonlinear manner.

Previous theoretical work[10] has also studied the effects of multisite phosphorylation on substrate rebinding. This model also predicts an exponential decay in the escape probability as a function of the number of phosphorylations. This effect then gives rise to a highly cooperative dose response curve (the addition of a single independent binding site can result in a large increase in the fraction of ligands bound to their receptors). However, a high degree of cooperativity is not observed in both our theoretical treatment and Monte Carlo simulations.

Although rebinding may not, in itself, produce a 'switch-like' (i.e. highly cooperative) dose response curve[2] in the fraction of ligands bound, it is nevertheless interesting to speculate on the ways in which the rebinding of a substrate to its receptor may affect myriad cellular processes. For instance, by controlling the probability of rebinding in the form of changing the number of phosphorylations on an enzyme, the degree of processive vs. distributive enzymatic modifications[2,28] that comprise a multi-step pathway could be controlled. It is also possible that the parameter $\gamma$ in our model that is determined by the relative size of the receptor and ligand and other structural features of the protein-protein interaction would be a key determinant in the number of processive versus distributive phosphorylation events[28].

Many mechanisms have been proposed (and some tested) that can account for switch-like dose responses involving proteins with multisite phosphorylations[15,22,29]. In the language of our model, such effects would result in $\theta$ having a complex, nonlinear relationship with $n$ and $\theta_0$. It may be interesting to explore how these mechanisms containing phenomena such as decoy phosphorylation, entropically driven binding, or electrostatics may couple to the effects of ligand rebinding as studied here.

Finally, the explicit geometry of the binding sites was not considered in this work. Other theoretical works[19,30] have shown that these effects can be important in polyvalent ligand binding. In future work, it may be interesting to investigate these geometrical aspects of multisite phosphorylation and ligand rebinding. Such a study might be accomplished, for example, by borrowing ideas from polymer physics[31], and considering the dynamics of a flexible polyvalent chain and its interaction with a substrate.

## Acknowledgements

This work was supported by an NIH Director's Pioneer Award granted to Arup K. Chakraborty and I gratefully acknowledge Arup K. Chakraborty for his support. I thank Jim Ferrell for introducing me to this problem. I am grateful to Steve Presse for a critical reading of the manuscript and his helpful comments.

## Appendix

### First passage time statistics and rebinding in three dimensions

The rebinding problem is now considered in three dimensions. Assuming spherical symmetry, the solution to the first-passage problem can be obtained in terms of modified Bessel functions. We introduce the survival probability

$$\Phi(\eta;t) = \int_t^\infty dt' F(\eta,t') = 1 - \int_0^t dt' F(\eta,t') \qquad \text{(A1)}$$

so that,

$$F(\eta;t) = -\frac{d\Phi(\eta,t)}{dt}. \qquad \text{(A2)}$$

In the Laplace domain:

$$\tilde{\Phi}(\eta;s) = \int_0^\infty dt e^{-st} \Phi(\eta;t), \qquad \text{(A3)}$$

The first passage time PDF can be written as follows:

$$\tilde{F}(\eta;s) = \Phi(\eta,0) - s\,\tilde{\Phi}(\eta,s) = 1 - s\,\tilde{\Phi}(\eta,s) \qquad (A4)$$

where, $\Phi(\eta,0) = 1$ (the survival probability at time zero is defined as 1). The survival probability can be obtained by solving a backwards Kolmogorov equation[25,32] that has the form of a diffusion equation

$$\frac{\partial \Phi(\delta,t)}{\partial t} = D\nabla^2 \Phi(\delta,t),$$

or, in the Laplace domain:

$$D\nabla^2 \tilde{\Phi}(\delta,s) = s\,\tilde{\Phi}(\delta,s) - 1 \quad, \qquad (A5)$$

(from hereon, length is scaled with respect to a diffusion length scale;

$\eta = \delta\sqrt{\dfrac{s}{D}}$ ; $\delta = a + \varepsilon$ ) with absorbing boundary condition,

$$\Phi(\eta_a,s) = 0 \qquad (A6)$$

where $\eta_a = a\sqrt{\dfrac{s}{D}}$ and $a$ is the radius of the sphere containing the targeted substrate.

For the other boundary condition, far away from the target at a distance, $\eta_0$, $\Phi(\eta;t)$ is unity; i.e.

$$\tilde{\Phi}(\eta \to \eta_0;s) = \frac{1}{s}. \qquad (A7)$$

In spherical coordinates, eq. A5 becomes:

$$\frac{d^2 \tilde{\Phi}(\eta;s)}{d\eta^2} + \frac{2}{\eta}\frac{d\tilde{\Phi}(\eta;s)}{d\eta} - \tilde{\Phi}(\eta;s) + \frac{1}{s} = 0 \qquad (A8)$$

and has the general solution:

$$\tilde{\Phi}(\eta; s) = \frac{1}{s} + A \frac{I_{-1/2}(\eta)}{\eta^{1/2}} + B \frac{I_{1/2}(\eta)}{\eta^{1/2}}$$

$$= \frac{1}{s} + \sqrt{\frac{2}{\pi}} \left[ A \frac{cosh(\eta)}{\eta} + B \frac{sinh(\eta)}{\eta} \right]. \qquad (A9)$$

where $I_v(x)$ is a modified Bessel function of order $v$.

The solution for $\tilde{\Phi}(\eta; s)$ that satisfies the boundary conditions in eqs. A6 and A7 gives the coefficients A and B:

$$A = \sqrt{\frac{\pi}{2}} \left( \frac{\eta_a}{s} \right) \frac{sinh(\eta_0)}{sinh(\eta_a - \eta_0)}$$

and

$$B = -\sqrt{\frac{\pi}{2}} \left( \frac{\eta_a}{s} \right) \frac{cosh(\eta_0)}{sinh(\eta_a - \eta_0)}.$$

Substituting the coefficients into eq. A9 and making use of the appropriate trigonometric identities gives:

$$\tilde{\Phi}(\eta; s) = \frac{1}{s} \left[ 1 - \frac{\eta_a sinh(\eta_0 - \eta)}{\eta sinh(\eta_0 - \eta_a)} \right] \qquad (A10)$$

Now we assume that the length of the total system (i.e. the cell) is much larger than the length of a single protein ($\eta_0 >> \eta_a$); so that $sinh(\eta_0 - \eta_a) \approx sinh(\eta_0)$ and $tanh(\eta_0 - \eta_a) \approx 1$. Upon substituting these relations and performing some algebraic manipulations, we obtain:

$$\tilde{\Phi}(\eta; s) \approx \frac{1}{s} \left[ 1 - \frac{\eta_a}{\eta} \left\{ sinh(\eta) - cosh(\eta) \right\} \right]$$

$$\approx \frac{1}{s}\left[1 + \frac{\eta_a e^{-\eta}}{\eta}\right] \tag{A11}$$

Substituting eq. A11 into eq. A4 gives an expression for the first passage time, $\tilde{F}(\eta;s)$:

$$\tilde{F}(a+\varepsilon;s) \approx \left[\frac{a}{a+\varepsilon} e^{-\sqrt{\tau s}}\right] \tag{A12}$$

where the distance $\delta$ is written as $\delta = a + \varepsilon$ and the variable $\tau = \frac{(a+\varepsilon)^2}{D}$ has been introduced. As in a prior study[15], a further simplification can be made if we observe the system on time scales commensurate with signaling times (times over which signals are propagated); $t \sim \frac{1}{k_-}$, i.e. $t \gg \tau$ (so that s is small); then $\tilde{F}(a+\varepsilon;s)$ becomes

$$\tilde{F}(a+\varepsilon,s) \approx \frac{a}{a+\varepsilon}\left[1 - \sqrt{\tau s}\right] + O(\tau s). \tag{A13}$$

Therefore, up to order $O(\tau s)$, we substitute eq. A13 into eq. 7 and obtain:

$$1 - \tilde{R}(a+\varepsilon,s) \approx \frac{(1-\gamma) + \gamma\sqrt{\tau s}}{1 - (1-\theta)\gamma} \tag{A14}$$

where $\gamma = \frac{a}{a+\varepsilon}$.

**First passage time statistics and rebinding in one dimension**

Although eq. 9 in 1d is exact, $\tilde{F}(\delta,s)$, however, often has a complicated form. Such a complication can make the Laplace inversion very difficult. For instance in the continuum limit in one dimension[25]:

$$F(\delta;t) = \frac{\delta}{(4\pi Dt)^{1/2}} \frac{e^{-\delta^2/4Dt}}{t} \tag{B1}$$

which has the Laplace transform: $F(\delta;s) = e^{-\sqrt{\alpha s}}$ -- $\alpha$ is the microscopic time scale that it takes a protein with diffusion constant D to diffuse a tiny amount, $\delta$; $\alpha = \dfrac{\delta^2}{4D}$ .

Subsequently, eq. 11 can be substituted into eq. 9 to obtain:

$$\tilde{R}\left(\delta,s\right) = \frac{\theta e^{-2\sqrt{\alpha s}}}{1 - \left(1-\theta\right)e^{-2\sqrt{\alpha s}}} \qquad (B2)$$

Despite this complication, additional simplifications can be made if we consider an observable time scale of signal transduction, $\tau_{sig} \sim (1/k_-)$, that is much longer than the microscopic diffusion time ($\alpha << \tau_{sig}$). In this case: $F(\delta;s) = e^{-\sqrt{\alpha s}} \approx 1 - \sqrt{\alpha s} + O(\alpha s)$.

So that upon substituting into eq. B2, we obtain:

$$R(\delta;s) \approx 1 - \frac{2\sqrt{\alpha s}}{\theta} . \qquad (B3)$$

As in a previous study[15], substituting eq. B3 into eq. 5 gives:

$$\tilde{f}\left(s\right) = \frac{f\left(0\right)}{s + \left(2k_-\theta\right)\sqrt{\delta s}} . \qquad (B4)$$

eq. B4 can be inverted[33]:

$$f\left(t\right) = f\left(0\right)e^{\kappa t}erfc\left(\sqrt{\kappa t}\right) \qquad (B5)$$

where $1/\kappa$ is a single characteristic time-scale ($\kappa = \dfrac{4k_-^2\alpha}{\theta^2}$) .

Figure 1.) **How biological responses might be shaped by allovalency, multisite phosphorylation, and ligand rebinding.**

A schematic for a ligand, with multiple equivalent binding sites, potentially rebinding to its enzyme. Once the ligand unbinds from its target, two possible outcomes are available: 1.) escape from its binding partner (i.e. diffuse a distance far away from the receptor) and 2.) immediate rebinding to its receptor. A biological response can then be initiated if the protein is bound sufficiently long. The outcome is expected to depend on the number of sites (in the form of phosphorylations of the protein) that are available. Circles depict different potential binding sites that arise from phosphorylations.

Figure 2.) **Strong ligand rebinding can be significant over many time scales.**

a.) Plots of the absorption probability in the Laplace domain, $R(\delta; s)$, with units chosen so that the microscopic diffusion time scale $\tau$ is unity, $\tau = 1$; ($\tau = \dfrac{(a+\varepsilon)^2}{D}$), are shown on a log-log plot. The strong rebinding limit is considered, $\gamma \to 1$, for convenience.

$R(\delta; t) = \dfrac{1}{2\pi i} \displaystyle\int_{c-i\infty}^{c+i\infty} ds R(\delta; s) e^{st}$ is the probability that a ligand absorbs to its target a distance

$\delta$ away at time $t$. $R(\delta; t)$ contains all known information on the statistics of an

individual ligand's past history of rebinding attempts. Plots are generated from the

expression obtained using eq. B2 . b.) Shapes of the dissociation curves in three limits:

1.) when no binding occurs, 2.) when $\gamma \to 1$ (strong rebinding), and 3.) when $0 < \gamma < 1$ or

$\varepsilon = O(a + \varepsilon)$ (weak rebinding). Dashed lines show the behavior of the decay curve in the

absence of rebinding, $k_- = 1$. Dotted lines give the case when the decay curve for rebinding takes the form of the strongly non-exponential one-dimensional case i.e. $\gamma = 1$.

The time constant, $\kappa$ ($\kappa = \dfrac{4k_-^2 \alpha}{\theta^2}$), in the appendix is taken to be unity $\kappa = 1$. Dash-

dotted lines show the behavior of the decay curve in the instance of weak rebinding limit

($k_-^{eff}(1-\gamma) = \dfrac{1}{2}$).

Figure 3.) **Rebinding is influenced by an increase in the number of phosphorylation sites.**

$f\left(\theta,t_0\right)$ is plotted for different values of $t_0$ given on the legend: the two regimes a.) strong rebinding, $\gamma \rightarrow 1$ and b.) weak rebinding regime, e.g. $\gamma = 0.9$; for both instances, $\tau = 10^{-6}s$, $k_- = 1s^{-1}$. $f\left(\theta,t_0\right)$ gives the probability that a ligand remains bound to its target as a function of the number of phosphorylation sites, $\theta$, and at a given time $t_0$.

When the time $t_0$ is commensurate with or greater than the intrinsic time constant $\dfrac{1}{k_-}$, i.e. $t_0 > \dfrac{1}{k_-}$, the positive contribution to the function, $f\left(\theta,t_0\right)$ is mostly due to rebinding.

Figure 4.) **Monte Carlo simulations suggest that an exponential decrease in the escape probability for an increasing number of phosphorylation sites can be insufficient to produce a switch-like dose response.**

Plots of simulation data from Monte Carlo simulations are shown. a.) The escape probability, $Pesc$ (defined in the methods section), as a function of $\theta$ ($\theta \sim e^{-\left(\frac{E_{k_b}}{k_b T}\right)}$, $\theta = n\theta_0$) is given. Three different values of the effective diffusion constant $D_{eff} \equiv \frac{P_{diff}}{P_{rxn}}$ are shown: $D_{eff} = 1$ (squares), $D_{eff} = 10$ (circles), and $D_{eff} = 100$ (crosses). The plot in the inset contains a fit to an exponential function $Pesc = e^{-k_{esc}\theta}$ for the $D_{eff} = 1$ case; $k_{esc} = 10^3 mcsteps^{-1}$ was used in the plot. b.) The fraction of bound ligand as a function of $\theta$ is shown. Four values of a scaled receptor density $\frac{\rho}{\rho_0}$, where $\rho_0 = 1000 receptors/cell$, are considered: $\rho = 1$ (squares), $\rho = 2$ (diamonds) $\rho = 5$ (circles) $\rho = 10$ (crosses). The plot in the inset gives a fit to a Hill function, $\frac{\theta^H}{K_{50\%}^H + \theta^H}$ with H = 1, for the case of $\rho = 1$. Error bars from the simulations are on the order of 5% of the reported values.

Figure 5.) **Graded responses are observed for over wide ranges of parameter values.**
Plots of $f\left(n, t_0\right)$ in which $\theta = n\theta_0$ are shown for different values of $\theta_0$. The number of phosphorylations, $n$ is plotted along the abscissa. Strong (a,b) and weak (c,d) rebinding limits are considered. Numbers on the legend indicate the different values of $\theta_0$ that were used. In the strong rebinding cases (a,b), two time points, $t_0$, are given: a.) $t_0 = 100s$ and b.) $t_0 = 1s$. In the weak rebinding cases (c,d), the two values of $t_0$ used were: c.) $t_0 = 5s$ and d.) $t_0 = 1s$.

Figure 1.)

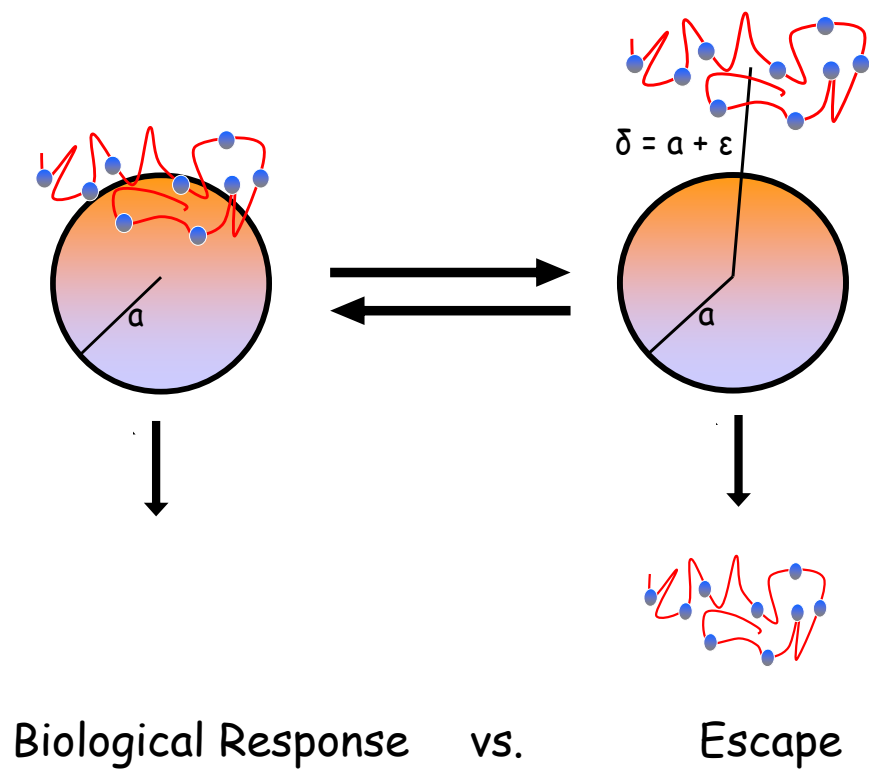

$\delta = a + \varepsilon$

Biological Response    vs.    Escape

Figure 2.)

a.)

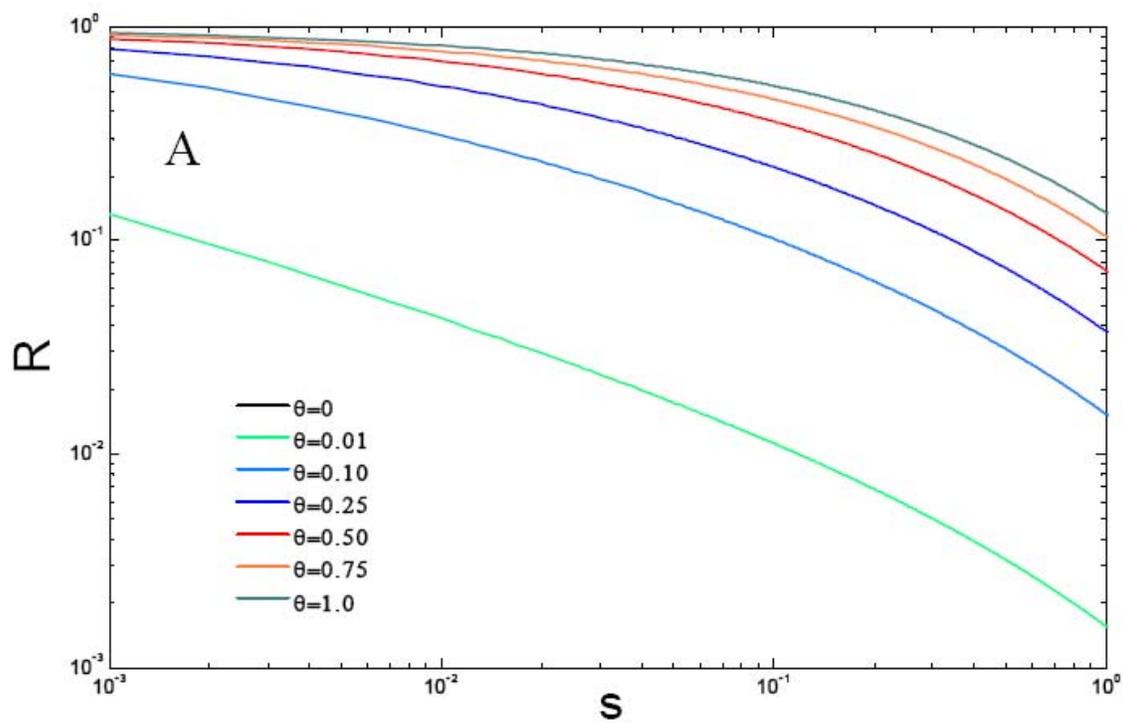

b.)

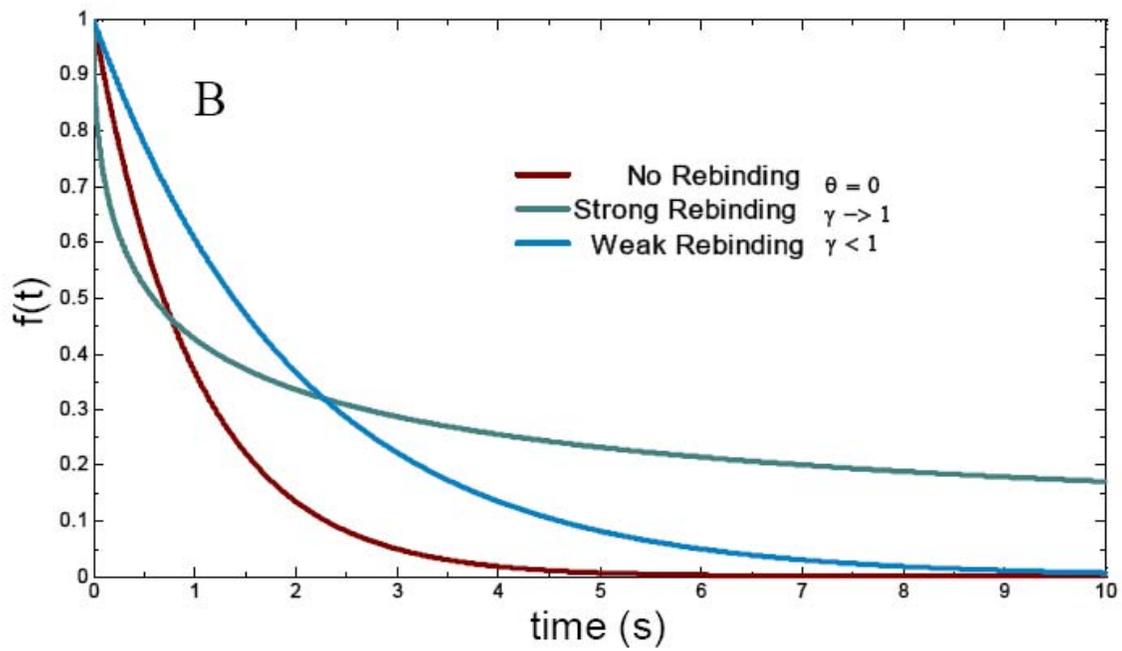

Figure 3.)

a.)

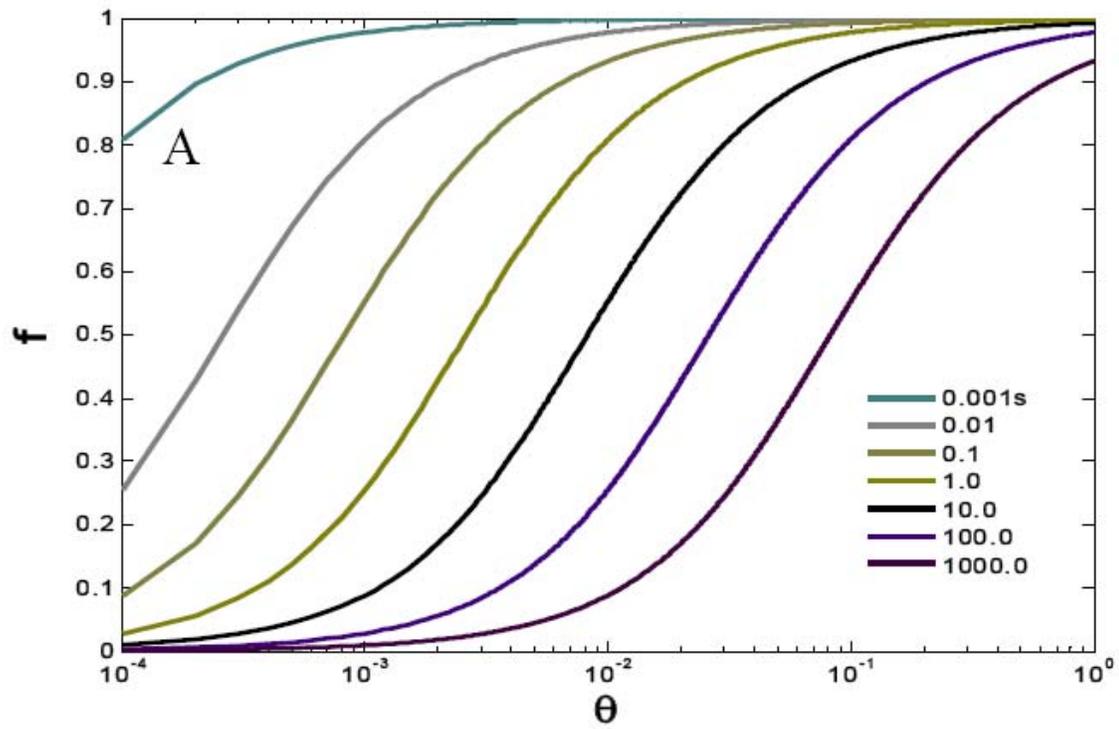

b.)

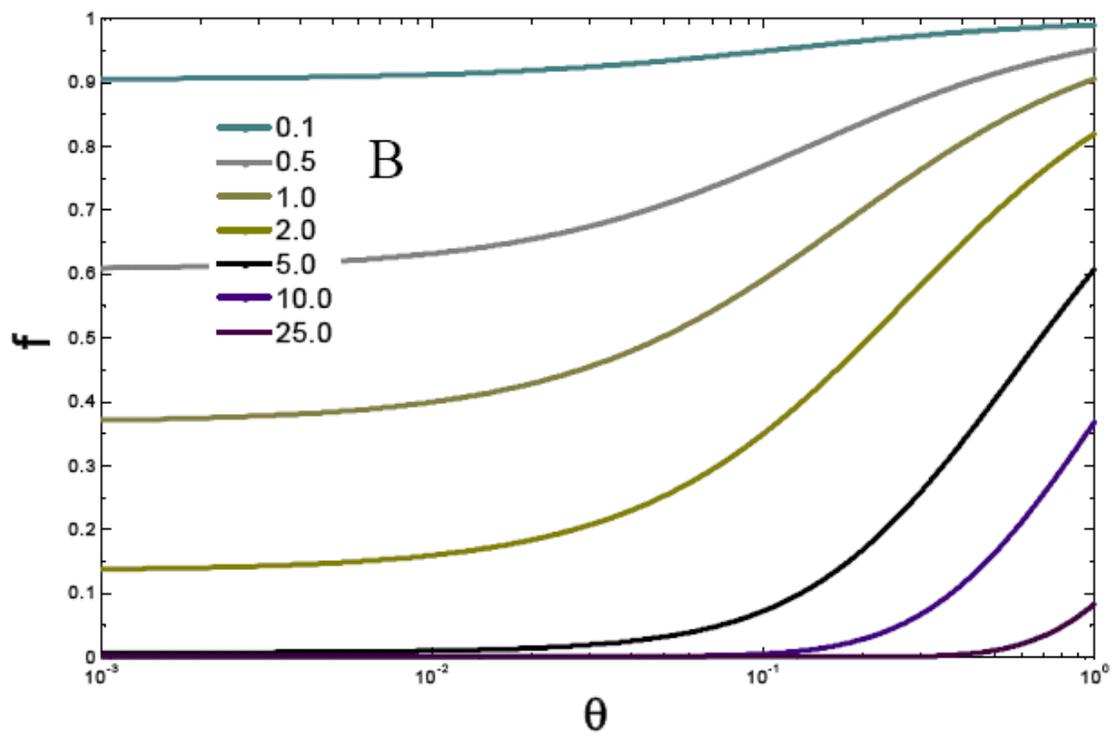

Figure 4.)

a.)

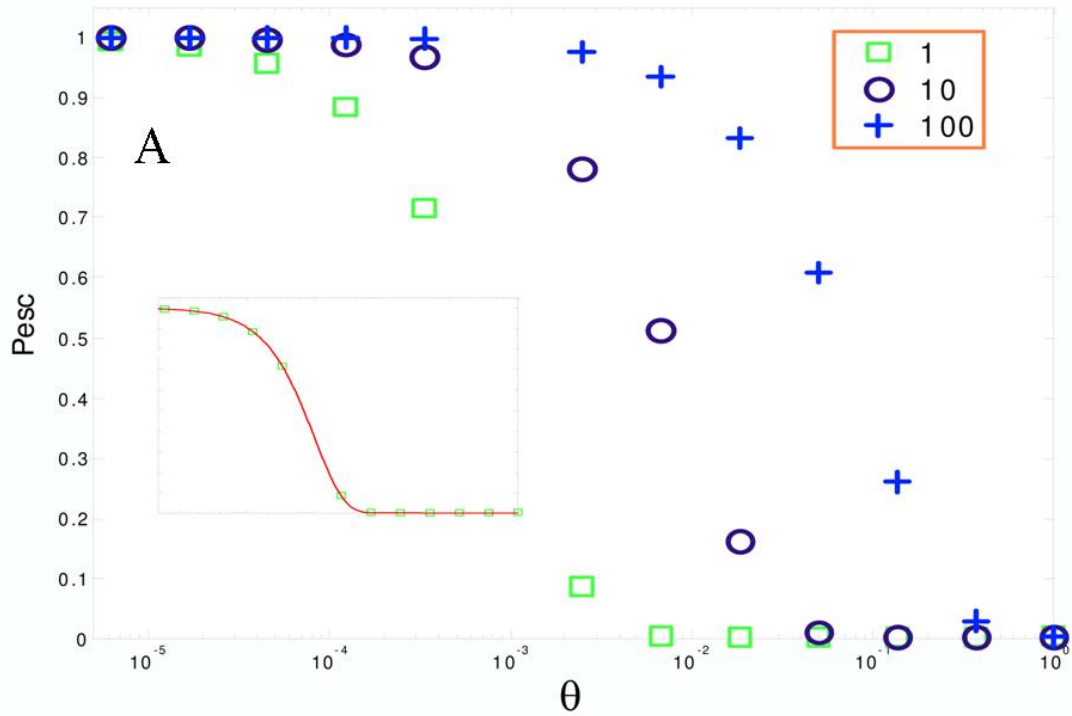

b.)

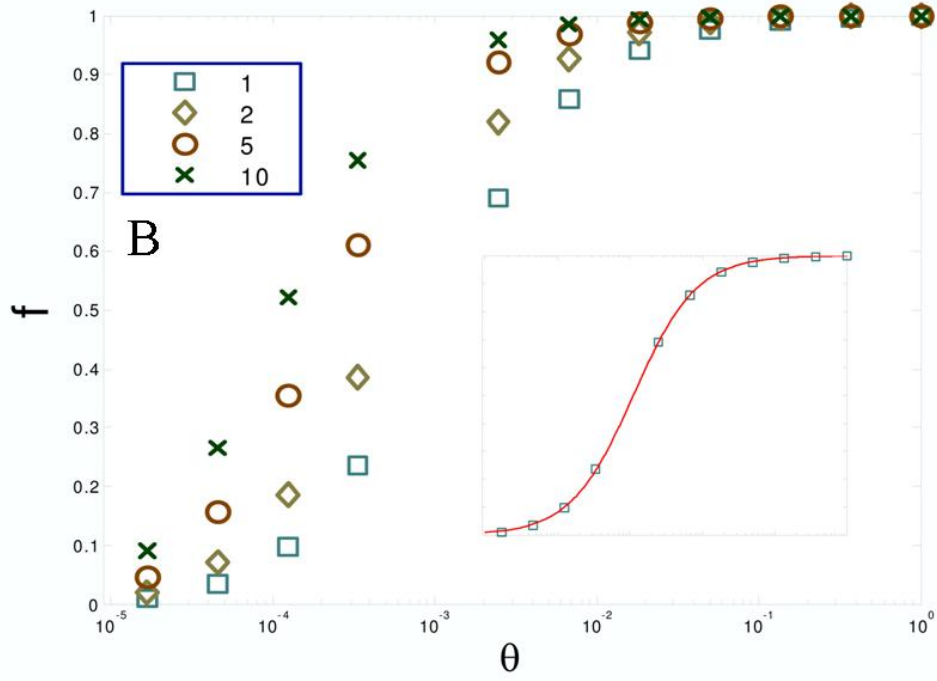

Figure 5.

a.)

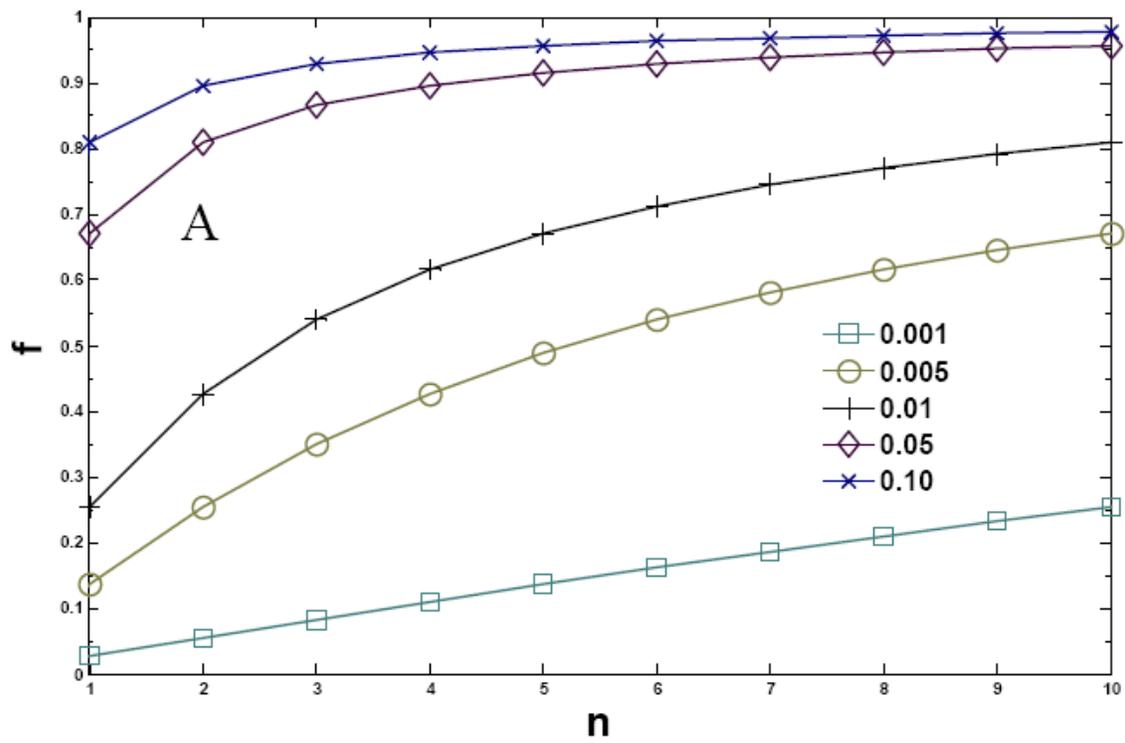

b.)

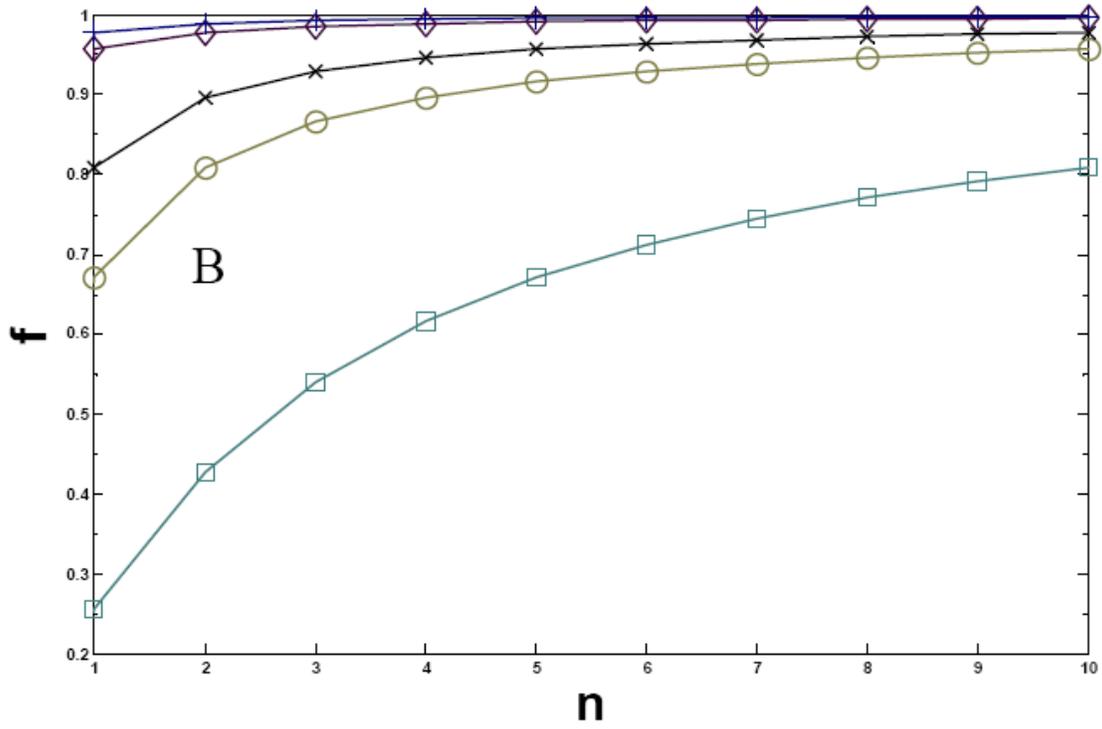

c.)

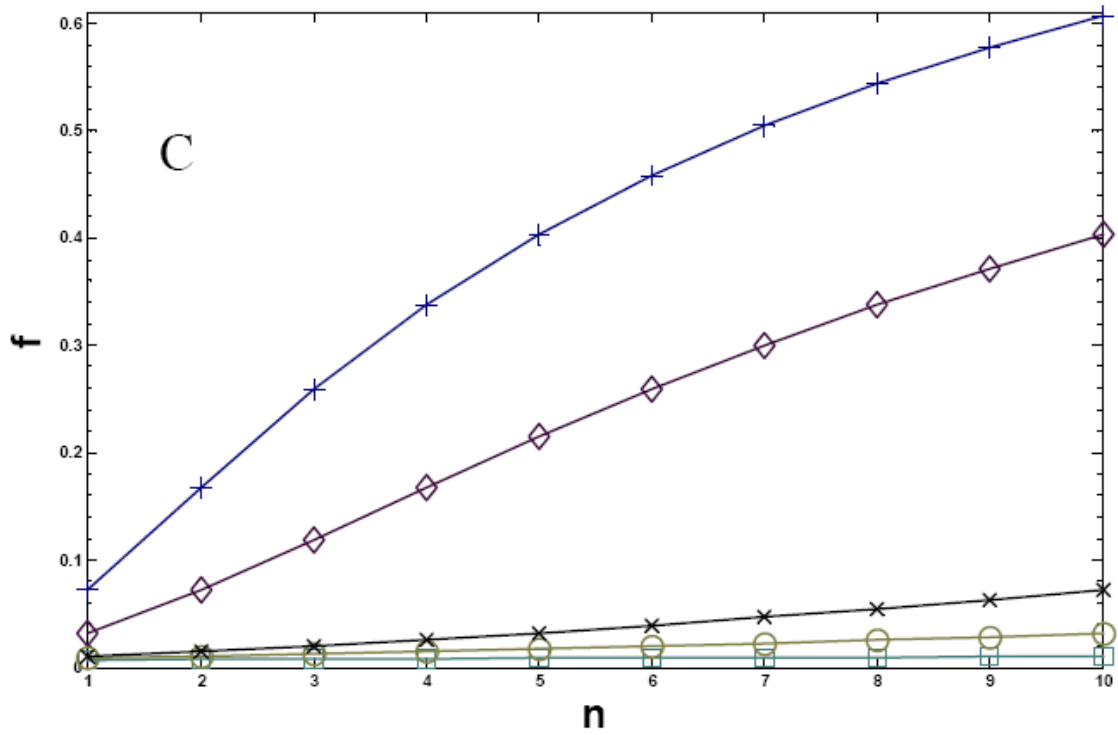

d.)

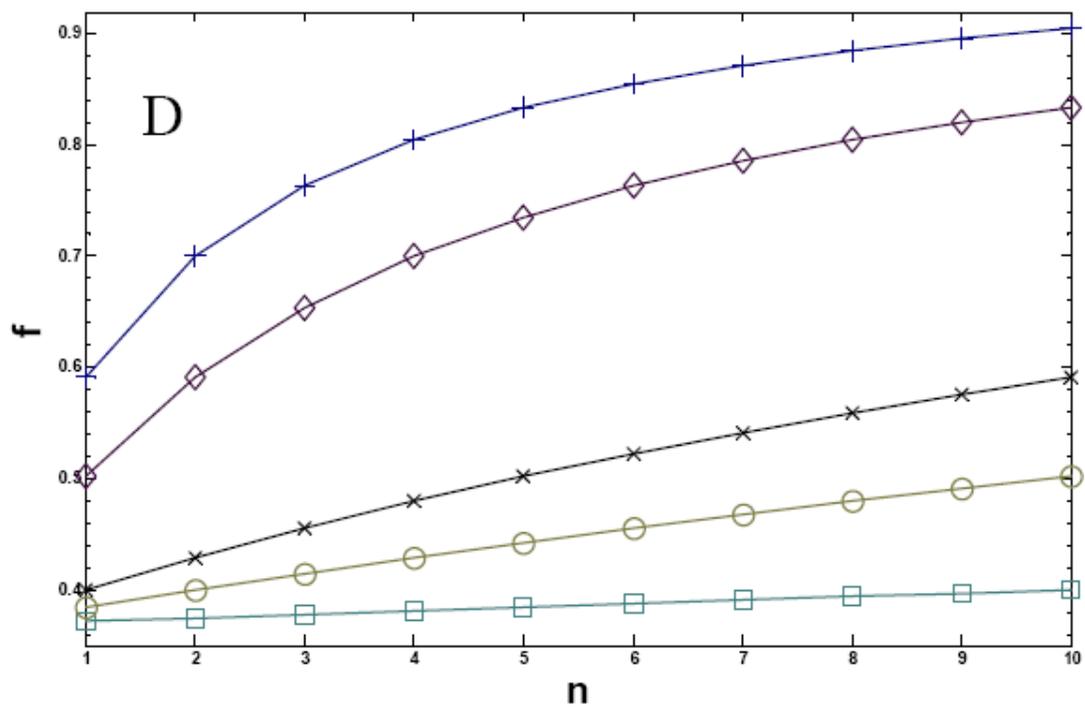